\documentclass[iop,apj]{emulateapj} \usepackage{apjfonts} \newif\ifdraft \drafttrue \newif\ifpre \pretrue
\usepackage{graphicx}
\usepackage{xcolor}
\ifdraft
   \usepackage{hyperref}
   \hypersetup{
               colorlinks=true,
               linkcolor=black,
               filecolor=black,
               citecolor=black,
               urlcolor=blue
              }
     \newcommand{\web}[1]{\Blb{\url{#1}}}
\else
     \newcommand{\web}[1]{#1}
\fi

\bibpunct{(}{)}{;}{a}{}{,}
\newcommand{\Frac}[2]{\frac{\displaystyle\strut #1}{\displaystyle\strut #2} }

\newcommand{\PIMA}{$\cal P\hspace{-0.067em}I\hspace{-0.067em}M\hspace{-0.067em}A$ }
\newcommand{\Number}[1]{\ifnum#1<10\relax0\number#1\else\number#1\fi}
\newcommand{\isodate}{
\count151=\time
\divide\count151 by 60
\count151=\count151
\multiply\count151 by 60
\count152=\time
\advance\count152 by -\count151
\divide\count151 by 60
\count152=\count151
\multiply\count151 by 60
\count153=\time
\advance\count153 by -\count151
\Number{\year}.\Number{\month}.\Number{\day}--\Number{\count152}:\Number{\count153}
}
\renewcommand{\aa}{A\&A}
\newcommand{\ntab}[2]{ \multicolumn{1}{#1}{#2} }
\newcommand{\nntab}[2]{ \multicolumn{2}{#1}{#2} }
\newcommand{\nnntab}[2]{ \multicolumn{3}{#1}{#2} }

\definecolor{Dred}{rgb}{0.312,0.070,0.070}
\definecolor{Dblue}{rgb}{0.070,0.070,0.312}
\definecolor{Dgreen}{rgb}{0.070,0.312,0.070}
\definecolor{Db}{rgb}    {0.050,0.0,0.320}

\newcommand{\Blb}[1]{\textcolor{Dblue}{\bf #1}}

\newcounter{note}
\let\oldmarginpar\marginpar
\renewcommand\marginpar[1]{\-\oldmarginpar[\raggedleft\footnotesize #1]%
{\raggedright\footnotesize #1}}

\newcommand{\hpm}{\hphantom{-}}
\ifdraft
\fi
\newcommand{\Gaia}{\emph{Gaia}}

\ifdraft
\fi
\begin{document}

\shorttitle{The position catalogue OBRS--2}
\title{The catalogue of positions of optically bright extragalactic
       radio sources OBRS--2}
\author{ L. Petrov}
\affil{Astrogeo Center, Falls Church, VA 22043, USA}
\email{Leonid.Petrov@lpetrov.net}
\ifdraft
  \journalinfo{Astronomical Journal}
  \submitted{}
  \received{2013, January 23}
  \accepted{}
  \published{}
\fi

\begin{abstract}

  It is anticipated that future space-born missions, such as \Gaia, will be
able to determine in optical domain positions of more than 100,\ 000 bright 
quasars with sub-mas accuracies that are comparable to very long baseline 
interferometry (VLBI) accuracies. Comparisons of coordinate systems from 
space-born missions and from VLBI will be very important, first for 
investigation of possible systematic errors, second for investigation 
of possible shift between centroids of radio and optical
emissions in active galaxy nuclea. In order to make such a comparison more
robust, a program of densification of the grid of radio sources detectable
with both VLBI and \Gaia\ was launched in 2006. In the second observing
campaign a set of 290 objects from the list of 398 compact extragalactic radio 
sources with declinations $>-10\degr$ was observed with the VLBA+EVN in 
2010--2011 with the primary goal of producing their images with milliarcsecond 
resolution. These sources are brighter than 18 magnitude at V~band. In this 
paper coordinates of observed sources have been derived with milliarcsecond 
accuracies from analysis of these VLBI observations following the method 
of absolute astrometry and their images were produced. The catalogue 
of positions of 295 target sources and estimates of their correlated flux
densities at 2.2 and 8.4~GHz is presented. The accuracies of source
coordinates are in a range of 2 to 200~mas, with the median 3.2~mas.

% It is anticipated that future space-born missions, such as Gaia, will be able to determine in optical domain positions of more than 100,000 bright quasars with sub-mas accuracies that are comparable to very long baseline interferometry (VLBI) accuracies. Comparisons of coordinate systems from space-born missions and from VLBI will be very important, first for investigation of possible systematic errors, second for investigation of possible shift between centroids of radio and optical emissions in active galaxy nuclea. In order to make such a comparison more robust, a program of densification of the grid of radio sources detectable with both VLBI and Gaia was launched in 2006. In the second observing campaign a set of 290 objects from the list of 398 compact extragalactic radio sources with declinations greater -10 deg was observed with the VLBA+EVN in 2010-2011 with the primary goal of producing their images with milliarcsecond resolution. These sources are brighter than 18 magnitude at V band. In this paper coordinates of observed sources have been derived with milliarcsecond accuracies from analysis of these VLBI observations following the method of absolute astrometry and their images were produced. The catalogue of positions of 295 target sources and estimates of their correlated flux densities at 2.2 and 8.4 GHz is presented. The accuracies of source coordinates are in a range of 2 to 200 mas, with the median 3.2 mas.

\end{abstract}

\keywords{astrometry --- catalogues --- surveys}

\maketitle

\section{Introduction}

   The method of Very Long Baseline Interferometry (VLBI) first proposed
by \citet{r:mat65} allows us to derive source positions with
nanoradian precision (1 nrad $\approx$ 0.2~mas). Since 1971 when the 
first catalogue of source coordinates determined with VLBI was published 
\citep{r:first-cat}, the number of extragalactic compact radio sources 
which positions were derived using VLBI under absolute astrometry 
observing programs grew from 35 objects to 7215 in 
2012\footnote{see \web{http://astrogeo.org/rfc}.}. For 95\% these
sources, accuracies of their positions are in a range of 0.05 to 6.5~mas 
with the median 0.5~mas. These sources form a dense grid on the sky
that can be used for many applications, such as differential astrometry,
phase-referencing VLBI observations of weak objects, space navigation,
Earth orientation parameter determination, and space geodesy.

  However, high accuracy of positions of these objects can be exploited
{\it directly} only by applications that utilize the VLBI technique.
Applications that use different observational techniques can benefit from
the high accuracy of VLBI positions only {\it indirectly} by observing common
objects from the VLBI catalogue with instruments at other wavelengths.
The European Space Agency space-born astrometry mission \Gaia, scheduled 
to be launched in 2013, according to \citet{r:gaia} promises to reach 
sub-mas accuracies of determining positions of quasars of 16--20 magnitude 
that will rival accuracies of absolute astrometry VLBI. Since position 
catalogues produced with \Gaia\ and VLBI will be completely independent, 
their mutual rotations, zonal differences and possibly other systematic 
effects can be interpreted as errors of one of the techniques after resolving 
the differences due to a misalignment of centers of optic and radio images 
of quasars and a frequency-dependent core-shift 
\citep{r:kov08,r:por09,r:sokol11}. Investigation of systematic differences 
will be very important for the assessment of the overall quality of \Gaia\
results and, possibly, the errors in the VLBI position catalogue.

  This comparison will produce valuable results if 1)~it will be
limited to those common sources which VLBI positions are known with errors
smaller than several tenths of a milliarcsecond; 2)~the number of sources
will be large enough to derive meaningful statistics; and 3)~the sources
will be uniformly distributed over the sky. However, the number of quasars
that have a compact core and are bright in both optical and radio 
wavelengths, and therefore, can be detected with both techniques, 
currently is rather limited. The observing program for densification of 
the list of such objects was launched in 2006 \citep{r:bou08} with 
the eventual goal of deriving highly accurate position of sufficiently 
radio-loud quasars from VLBI observations in the absolute astrometry mode. 
The original observing sample consisted of 447 optically bright, relatively 
weak extragalactic radio sources with declinations above $-10^{\circ}$. 
The detailed observing scheme of this project is presented in \cite{r:bou08}. 
The first VLBI observing campaign in 2007 resulted in detection of 
398 targets with the European VLBI Network (EVN) \citep{r:bou10}, 
although no attempt to derive their positions or produce images was 
made. During the second observing campaign a subset of 105 sources 
detected in the previous campaign was observed with the global VLBI 
network that comprises the VLBA and EVN observing stations with the goal 
of revealing their morphology on milliarcsecond scales from VLBI 
images \citep{r:bou11} for consecutive screening the objects with 
structure that potentially may cause non-negligible systematic 
position errors. Their positions were derived by \citet{r:obrs1} and 
formed the OBRS--1 catalogue.

  In 2010--2011 remaining 290 sources have been observed in the third
campaign, hereafter called OBRS-2, with the global network that comprises 
the VLBA and EVN observing stations in a mode similar to the second 
campaign. I present here results of data analysis of this observations. 
Observations and their analysis are described in sections \ref{s:obs} 
and \ref{s:anal}. The position catalogue is presented in section 
\ref{s:cat} and discussed in section \ref{s:discussion}. Concluding 
remarks are given in section \ref{s:summ}.

\section{Observations}
\label{s:obs}

   During OBRS--2 campaign there were three observing sessions with 10 VLBA 
stations and 5--6 EVN stations from this list: {\sc eflsberg}, {\sc medicina}, 
{\sc onsala60}, {\sc yebes40m}, {\sc dss63}, {\sc hartrao}, {\sc noto}. 
First four EVN stations participated in every experiment, three remaining 
stations participated in some experiments. Each program source was observed 
in one session, in 3--4 scans, each 5 minutes long. Two sources,
1148$+$387 and 1203$+$109, were observed in two sessions in 5 scans.
In addition to 290 program sources, 8 strong calibrators were observed.
   
\begin{table}
    \caption{Summary of observing sessions}
    \begin{center}
       \begin{tabular}{lllr}
          2010.03.23 & gc034a   & 48 &  97 \\
          2011.11.08 & gc034bcd & 58 & 118 \\
          2011.03.15 & gc034ef  & 40 &  77 \\
       \end{tabular}
    \end{center}
    \tablecomments{The columns are: 1)~epoch of an observing session, 
                   2)~session ID, 3)~total session duration in hours 
                   excluding maintenance gaps, and 4)~the number of 
                   program sources.}
    \label{t:sess}
\end{table}

   Observations were made at X and S~bands simultaneously. 
Data were sampled with 2 bits per sample at the aggregate
rate of 512~Mbit/s. The intermediate frequencies were selected 
to provide continuous bandwidth [8.37699, 8.44099] and 
[2.22699, 2.29099]~GHz. Such a setup is not favorable for 
astrometry. Spanning frequencies over 500~MHz, as it done 
in all other dual-band absolute astrometry VLBA surveys, 
improves precision of group delay determination, and therefore, 
source position precision {\it by one order of magnitude}. 
The position accuracy of sources observed with a wide-band 
frequency setup and detected at baselines 3,000--8,000~km long
is limited by systematic errors. The position accuracy of sources
observed in a frequency band spanned over 64~MHz is limited by 
the thermal noise.

  Second limitation of the OBRS--2 schedule for the astrometry use
is a relatively rare observation of sources at low and high 
elevations for better estimation of the troposphere path delay 
in zenith direction, which increases systematic errors. 
VCS1 survey~\citep{r:vcs1} suffered a similar deficiency 
in design, and analysis revealed systematic errors at a level 
of 0.4--0.5~mas. We can assume systematic errors of OBRS--2 
were at a similar level. But since the contribution of the 
thermal noise in the error budget of OBRS--2 experiments is one 
order of magnitude higher than in VCS experiments, these increased 
systematic errors did not cause a noticeable further accuracy 
degradation.

  Of 290 program objects, 58 sources were observed and detected 
in other absolute astrometry programs, such as Very Long Baseline 
Array (VLBA) Calibrator Survey 
\citep{r:vcs1,r:vcs2,r:vcs3,r:vcs4,r:vcs5,r:vcs6}, regular VLBA 
geodetic observations \citep{r:rdv,r:pus12}, and ongoing VLBI 
observations of 2MASS galaxies \citep{r:v2m}. Positions of all
but four sources 0106$+$315, 0809$+$483, 1213$+$097, and 2247$+$140,
are known with accuracy better than 0.4~mas. These 54 program
sources and 8 calibrators provided a sufficient overlap for tying
position estimates to the existing catalogue and for evaluation of
position accuracy.

\section{Data analysis}
\label{s:anal}

  The data were correlated with DiFX software correlator \citep{r:difx2}
at the National Radio Astronomy observatory. The correlator computed 
the spectrum of cross correlation and autocorrelation functions with 
a frequency resolution of 0.25~MHz at accumulation intervals of 1~s long.

\subsection{Preliminary analysis}

  The procedure of further analysis is described in full detail
in \citet{r:vgaps}. I present here only a brief outline. First,
log files are parsed, system temperature and phase calibration
are analyzed, and points with outliers are removed. No phase 
calibration signal for non-VLBA stations was recovered due 
to a limitation of the DiFX correlator (which was lifted later 
in 2011). Next, the fringe amplitudes were corrected for
a distortion in the sampler due to digitization. Then, the amplitudes
were calibrated by multiplying them by the a~priori System Equivalent
Flux Density (SEFD) 
$\sqrt{ \Frac{g_1\, g_2}{T_{\rm sys1}\, T_{\rm sys2}}}$, where 
$g_i$ and $T_{\rm sysi}$ are the gain and system temperature of 
the $i$th antenna respectively. Then a group delay, phase delay rate, 
group delay rate, and fringe phase were determined for every 
observations for each baseline at X and S~bands separately using 
the wide-band fringe fitting procedure. These estimates maximize 
the amplitude of the cross-correlation spectrum coherently averaged 
over all accumulation periods of a scan and over all frequency channels 
in all intermediate frequency of a band. After the first run of fringe 
fitting, 12 observations at each baseline with a reference station 
with the strongest signal to noise ratios (SNRs) were used to adjust 
the station-based complex bandpass corrections, and the fringe fitting 
procedure with the bandpasses applied was repeated. This part of 
analysis is done with 
\PIMA\ software\footnote{Available at \web{http://astrogeo.org/pima}.}. 

  Further analysis was split into two routes: astrometric and imaging.
Following the astrometric route, total group delays and phase delay 
rates were computed on a common fringe reference time epoch within 
a scan using results of fringe fitting. These observables, along with 
auxiliary information describing observations, were exported to 
the VTD/post-Solve VLBI analysis software\footnote{Available 
at \web{http://astrogeo.org/vtd}.} for interactive analysis. Initially,
only observables with the high SNR to ensure that the probability of 
false detection is less than 0.001 were chosen. This SNR threshold 
is $5.8$ for OBRS--2 experiments. Detailed description of the method 
for evaluation of the probability of false detection can be found 
in \citep{r:vgaps}. 

  Then, theoretical path delays were computed according to the 
state-of-the art model as well as their partial derivatives. Small 
differences between group delays and theoretical path delay were used 
for estimation of corrections to a parametric model that describes the 
perturbation of the theoretical model using least squares (LSQ). 
Coordinates of target sources, positions of all stations, except the 
reference one, parameters of the spline with the time span of 1~hour 
that models corrections to the a~priori path delay in the neutral 
atmosphere in the zenith direction for all stations, and parameters 
of another spline with the same time span that describes the clock 
function for all stations but the reference one were solved for in 
separate least square solutions that used group delays at X and S~bands 
respectively.

   The dataset was cleaned for outliers, i.e. observations with residual
group delays exceeding $5\sigma$. The most common reasons for 
an observation to have a large residual are a failure of fringe fitting 
procedure to find the main maximum, the presence of radio interference, 
or false detection. Simultaneous solving for source positions with only 
few detections in a presence of outliers poses a certain risk. One bad 
observation may corrupt the solution, and as a result, remaining good 
observations of that source may get high residuals and the outlier 
elimination procedure may discard them. Identifying that bad 
observation(s) often required several trials. When the dataset was cleaned, 
I gradually lowered the SNR limit from 5.8 to 5.0 with a step of 0.1.  
The status of all observations with SNR $< 5.8$ was initially set to 
``suppressed''. That means these observations did not contribute 
to estimation of the parametric model, but the parametric model evaluated 
at the previous step was applied for computing their residuals. Suppressed 
observations with residuals by modulo less than $5\sigma$ at S~band and 
less than $4\sigma$ at X~band were restored one by one starting with 
the smallest normalized residual. After flipping the status ``suppressed'', 
the LSQ solution was updated for including this observation into the 
parameter estimation model, and the process was repeated. 
There was a significant fraction of false detections with SNR $< 5.8$. 
But such observations have random group delay estimates uniformly 
distributed over 4$\mu$s wide search window. The probability that 
the group delay from a false detection will have a residual within 
2--5~ns of the expected value, and therefore may be identified as 
a good observation, is an order of 1--$2 \cdot 10^{-3}$. Suppression 
of observations with huge residual delay rates (several such points 
were found) allows to reduce this probability even further.

  Then the fringe fitting procedure was repeated with a narrow fringe 
search window for those points that were marked as outliers. The center of
fringe search window over delay and delay rate was set to the expected
value of the delay and delay rate computed as a sum of the a~priori delay or
delay rate and the contribution from the parametric model derived during the 
LSQ adjustment. The width of the search window was set to 5~ns for S~band 
and 3~ns for X~band. In addition to that, I re-ran the fringe fitting 
procedure with the updated a~priori model for all the sources that have 
position adjustments exceeding $1''$. For instance, 0744$+$092 had 
the a~priori position $4.''8$ off the VLBI position. When a source has 
a large position error, a non-linear fringe phase change over a scan 
of 300~s long becomes significant and causes noticeable de-correlation.

  The interactive analysis procedure was repeated with updated results
of fringe fitting. The SNR threshold was lowered to 4.8 because 
the probability of false detection is less for a narrow fringe search 
window. The procedure of outliers elimination was repeated. Baseline 
dependent additive weight corrections were computed in such a way as 
the ratio of the weighted sum of post-fit residuals to their mathematical 
expectation was close to unity. This computation procedure is described
in full detail by \citet{r:rdv}. 

\subsection{Multiple sources}

  Careful analysis revealed five sources that had more than 15\% points 
with SNR $> 5.8$ marked as outliers that were not diagnosed as radio 
interference or errors in fringe fitting. These sources had different 
positions derived from X and S-band group delays. If to invert suppression 
status of points for these sources, i.e. to restore a point that was 
suppressed and suppress the point that was used in the solution, and to
re-run the procedure of outlier elimination then we can get a position 
that is consistent to a position at the opposite band. These can be 
explained if to suggest that a sources has multiple components separated 
at an arcsecond level, strong enough to be detected. I used the following 
technique for component separation in order to investigate these sources. 
I cloned visibilities of these sources and treated them as different 
objects with different positions. Using their preliminary positions 
as a~priori, a shifted phases of visibilities to the new positions and 
performed the fringe search in a narrow window, the same way as I treated 
outliers. This approach yielded a significant number of new detections 
for both components at one or both bands. New detections confirmed 
the hypothesis that a source is multiple. 

\subsection{Global astrometric solution}

  The result of the interactive solution provided a clean dataset of
X and S-band group delays with updated weights. The dataset that was used 
for the final parameter estimation utilized all dual-band S/X data acquired 
under the absolute astrometry and space geodesy programs from April 1980 
through December 2012, including 76079 observations from OBRS--2 experiments, 
in total 8.89 million observations. As I mentioned, among program sources,
58 common objects were observed in other absolute astrometry VLBI 
experiments. I made four solutions. The first three solutions used the 
global dataset, except observations of 58 common objects, and observables 
from the OBRS--2 experiments: 1)~the first solution used the X-band 
group delays, 2)~the second solution used the S-band group delays,
and 3)~the third solution used the ionosphere free linear combinations 
of X and S-band group delays. The fourth reference solution used all 
experiments in the global dataset with the only exception of OBRS--2 data.

  OBRS--2 experiments were analyzed exactly the same way as 5497 other 
VLBI observing sessions, using the same analysis strategy that was used 
for processing prior observations for ICRF \citep{r:icrf98}, VCS, VGaPS, 
Long Baseline Array Calibrator Survey (LCS) \citep{r:lcs1}, and K/Q survey 
\citep{r:kq} catalogues. The estimated parameters are right ascensions 
and declination of all sources, coordinates and velocities of all stations, 
coefficients of B-spline expansion of non-linear motion for 26 stations, 
coefficients of harmonic site position variations of 48 stations 
at 4 frequencies: annual, semi-annual, diurnal, semi-diurnal, 
and axis offsets for 69 stations. Estimated variables also included Earth 
orientation parameters for each observing session, parameters of clock 
function and residual atmosphere path delays in the zenith direction 
modeled with the linear B-spline with interval 60 and 20 minutes respectively.
All parameters were adjusted in a single LSQ run.

  The system of LSQ equations has an incomplete rank and defines a family
of solutions. In order to pick a specific element from this family, I applied
the no-net rotation constraints on the positions of 212~sources marked 
as ``defining'' in the ICRF catalogue that required the positions of these 
sources in the new catalogue to have no rotation with respect to their 
positions in the ICRF catalogue. No-net rotation and no-net-translation 
constraints on site positions and linear velocities were also applied. 
The specific choice of identifying constraints was made to preserve 
the continuity of the new catalogue with other VLBI solutions made during 
last 15 years.

\ifdraft \newpage \fi
\subsection{Image analysis}

  The same dataset of visibilities was used for source imaging. At first,
I discarded all visibilities from observations that were marked 
as outliers in the final step of the interactive analysis procedure.
Next, the data should be averaged over time and frequency after phase
rotation for the contribution of group delay, group delay rate, and phase
delay rate found by the fringe search procedure. Since the fringe fitting 
procedure was baseline-based, the baseline-dependent parameters of fringe 
fitting should be transformed to station-based parameters in order to 
preserve phase closures of visibilities. This transformation was done 
for the visibilities of each scan, each subarray, and each band 
separately. Although the experiments were scheduled for all the stations 
of the network to observe the same source during scan time, it may happen 
that some sources were detected only at a subset of baselines that do not 
have common stations. For each scan I found a scan-reference time as 
a weighted mean epoch of used observations. I selected a reference station 
for each subarray and solved with least squares for station-dependent 
group delays, group delay rates, and phase delays using baseline-dependent 
estimates of these quantities found by the fringe search procedure 
as the right-hand-side. These station-dependent quantities related to 
a common epoch within a scan were used for phase rotating the visibilities. 
After phase rotation, the visibilities were averaged over 32 spectral 
channels in each intermediate frequency and over 4~s intervals.

  The averaged visibilities and accompanying weights were split into sources
and written in separate files. Further processing was done using DIFMAP
software package \citep{r:difmap}. I used automatic imaging procedure 
mupet developed by Martin Shepherd and Greg Taylor. It started from 
a point source model as an initial guess and performed a sequence of
image cleaning, phase and amplitude self-calibrations with and without
taper. 

  I developed a web application that allowed me to inspect images visually
and flag those that showed visible artifacts. Of 564 images made 
automatically, I selected 63 images that I processed interactively. 
These were the sources that either had points with amplitude outliers
that an automatic procedure was unable to flag out or sources with too
few observations, in a range of 15--35, when an automatic procedure of 
hybrid imaging becomes unstable. In total, images were produced for 279 
sources at X~band and for 285 sources at S~band\footnote{Images 
in FITS format as well as calibrated visibilities are available at 
\web{http://astrogeo.org/obrs}.}. As a measure of a source 
strength, I derived two quantities from source brightness 
distributions: I computed the median correlated flux densities at
baselines with projection lengths shorter than 900~km and 
the median correlated flux densities for baselines with projection 
lengths longer than 5000~km.

  Some weak sources have too few points for successful imaging.
In order to provide a measure of source strength at long and short 
baselines for these objects, I made a simplified amplitude analysis
similar to what was done in processing OBRS--1 observations. I used 
averaged corrections to gains evaluated during self-calibration of 
strong sources and applied these corrections to the a~priori SEFDs. 
From fringe amplitudes calibrated this way I computed the median 
correlated flux densities at baselines with projection lengths shorter 
than 900~km and for baselines with projection lengths longer than 
5000~km, similar to what I did in image processing. Comparison of 
estimates of median correlated flux densities of strong sources derived 
by this method with estimates of correlated flux densities derived from 
images produced by a rigorous self-calibration procedure showed 
they agree at a level of 15\%.
  
  Detailed analysis of produced images goes beyond the scope of the 
present paper and will be given in the future in a separate publication 
(G.~Bourda et al., 2013, paper in preparation).

\section{Results}
\label{s:cat}

  I have detected at least at one band all but one source 0843$-$025  
(J0845$-$0241). In present paper objects with multiple components are 
treated as different sources, although most likely these are parts of 
the same objects.

  Since 58 of 295 detected sources have been observed with VLBA in
different programs at X~band in a wide-band mode with frequency 
spanned over 494 and 992~MHz, their position uncertainties are 
a factor of 10--20 better than the position uncertainties from OBRS--2
experiments with the same SNR and the same number of observables. 
Therefore, for the purpose of comparison with OBRS--2 catalogue, positions
of these sources can be considered as precisely known.  I excluded from
comparison 4 sources that are resolved and had uncertainties exceeding
0.4~mas in the reference solution.

  Comparison showed that the solution that used ionosphere free linear 
combinations of X/S group delay observables did not improve the agreement 
between position estimates of 54 common sources with respect to 
the solution that used X-band only observables. The position 
uncertainties of OBRS--2 observations are too large for the residual 
ionosphere contributions to affect positions at a significant level. 
I computed the variance that, being added in quadrature to source 
position uncertainties, makes the ratio of the sum of weighted squares
of position differences to their mathematical expectations close 
to unity. These variances are 1.7~mas in right ascension and 2.1~mas 
in declination for the X-band solution, and 2.1~mas in right ascension 
and 2.4~mas in declination for the S-band solution.

\subsection{OBRS--2 catalogue}

  The majority of sources were detected at both bands, and there are 
position estimates from two solutions that used X-band and S-band 
observables from the OBRS--2 experiments. The preference was given 
to the solution that used X-band observables because first, 
the detection limit at X~band was lower and second, the contribution 
of the residual ionosphere after applying reduction for the total 
electron contents from GPS observations at X~band is one order of 
magnitude less than at S~band. If the reweighted position uncertainty 
from the S-band solution was at least a factor of 1.5 smaller than 
the uncertainty from the X-band solution, the position from the S-band 
solution was used in the final catalogue. Although there are additional
more precise observations of 58 common sources in other absolute 
astrometry experiments, these observations were excluded in solutions 
1 and 2 used for deriving the OBRS--2 catalogue.

\begin{table*}[t]
   \caption{First 8 rows of the OBRS--2 source position catalogue.}
   \label{t:cat}
   \ifdraft \relax \else \tiny \hspace{-3em} \fi
   \begin{tabular}{ l l r r r r r @{\hspace{0.5em}} c @{\hspace{0.25em}} r @{\hspace{0.5em}} r r r r r r r}
      \hline
      \nntab{c}{IAU name} &
      \nntab{c}{Source coordinates} &
      \nnntab{c}{Position errors} &
       &
      \nntab{c}{\# pnt} &
      \nntab{c}{$F_{corr}$ S band} &
      \nntab{c}{$F_{corr}$ X band} &
      \nntab{c}{Flags} 
      \\
      B1950  &
      J2000  &
      \ntab{c}{$ \alpha $ } &
      \ntab{c}{$ \delta $ } &
      $ \sigma_\alpha  $ &
      $ \sigma_ \delta $ &
      \ntab{c}{corr} &
      band &
      S &
      X &
      short &
      unres &
      short &
      unres &
      \\
        & 
        &
        \ntab{l}{~hr~mn~sec} &
        \ntab{l}{~~~~$^{\circ}$~~~~$^\prime$~~~~~$^{\prime\prime}$}  &
        \ntab{c}{mas} & \ntab{c}{mas} & & &
        & &
        \ntab{c}{Jy}  & \ntab{c}{Jy}  &
        \ntab{c}{Jy}  & \ntab{c}{Jy}  &
        S & 
        X \\
      \hline
        J0001$+$0632 & 2358$+$062 &  00 01 23.694601 & $+$06 32 30.93754  &  3.52 & 15.24 & $-$0.811 &  X &  38 &  31 &  0.018 & $\hpm$0.022 & 0.018 & 0.008 & m & m \\
        J0005$+$1609 & 0003$+$158 &  00 05 59.237650 & $+$16 09 49.02157  &  1.86 &  2.53 & $-$0.342 &  X & 168 & 168 &  0.143 & $\hpm$0.070 & 0.196 & 0.057 & m & m \\
        J0015$+$3052 & 0012$+$305 &  00 15 36.022281 & $+$30 52 29.79522  &  2.98 &  7.35 & $-$0.437 &  X &  12 &  32 &  0.026 & $\hpm$0.015 & 0.017 & 0.006 &   & m \\
        J0017$+$1451 & 0015$+$145 &  00 17 36.903866 & $+$14 51 01.88067  &  1.92 &  3.10 & $-$0.557 &  X & 164 & 166 &  0.065 & $\hpm$0.039 & 0.053 & 0.030 & m & m \\
        J0027$+$4514 & 0025$+$449 &  00 27 42.262713 & $+$45 14 57.07879  &  2.72 &  2.88 & $-$0.477 &  X & 222 & 129 &  0.048 & $\hpm$0.042 & 0.024 & 0.015 & m & m \\
        J0035$+$1553 & 0033$+$156 &  00 35 55.537977 & $+$15 53 16.45642  & 23.81 & 23.11 & $-$0.267 &  S &  29 &   3 &  0.074 & $<   $0.017 &-1.00  &-1.00  & m &   \\
        J0037$+$3938 & 0034$+$393 &  00 37 36.725767 & $+$39 38 11.79014  &  2.44 &  2.86 & $-$0.354 &  X & 210 & 121 &  0.142 & $\hpm$0.044 & 0.039 & 0.012 & m & m \\
        J0041$-$0143 & 0038$-$019 &  00 41 26.008767 & $-$01 43 15.67847  &  2.03 &  5.16 & $-$0.625 &  X & 143 & 125 &  0.056 & $\hpm$0.049 & 0.037 & 0.032 & m & m \\
      \hline
   \end{tabular}
   \tablecomments{Units of right ascension are hours, minutes and seconds. 
                  Units of declination are degrees, minutes and seconds.
                 \hspace{10em}\hfill\linebreak
                 (This table is available in its entirety in machine-readable
                  and Virtual Observatory (VO) forms in the online journal.
                  A portion is shown here for guidance regarding its form
                  and content.)
                 }
\end{table*}

  The positions of 295 sources observed in OBRS--2 experiment are listed 
in Table~\ref{t:cat}. The 1st and 2nd columns provide the IVS source name 
(B1950 notation) and IAU name (J2000 notation). The 3rd and 4th columns 
give source coordinates at the equinox on the J2000 epoch. Columns 5 and 6 
give reweighted source position uncertainties in right ascension and 
declination in mas (without $\cos\delta$ factor), and column 7 gives 
the correlation coefficient between the errors in right ascension and 
declination. Column~8 shows band ID of the solution that was used to derive 
position of a given source. The number of group delays used in analysis 
is listed in columns 9 and 10. Columns 11 and 12 provide the median value 
of the correlated flux density in Jansky at S~band at baseline projection 
lengths shorter than 900~km and at baseline projection lengths longer than 
5000~km. The latter estimate serves as a measure of the correlated flux density 
of an unresolved source component. Columns 13 and 14 provide the median 
of the correlated flux density at X~band at baselines shorter than 900~km and 
longer than 5000~km. If no information about correlated flux density 
is available, $-1.000$ is used as a placeholder. The last two columns have 
flags whether an image is available for S and X~bands: ``m'' if available, 
blank if not.

  The semi-major axes of error ellipses range from 2.1 to 200~mas,
with the median 3.2~mas, and for 80\% sources the position errors
are under 5.2~mas. The major factor that results in position uncertainty 
exceeding 5~mas is a lack of detections at long baselines. Sources
with large position uncertainties are either highly resolved 
or very weak moderately resolved objects.

\subsection{Analysis of multiple sources}

{\bf 0154$+$31A/0154$+$31B} has two components $2''.592$ apart.
Component~A is compact and was detected at 5~mJy level at X~band 
at intercontinental baselines with Effelsberg. It was not detected
at intercontinental baselines at S~band. Since the sensitivity of 
baselines with Effelsberg is a factor of 4--5 worse at S~band
than at X~band, this can be explained if a source has a spectral 
index steeper than $-1.1$ ($S \sim f^{\alpha}$). Component~B is 
stronger at S~band, but has only three detections at 5~mJy
level at X~band at short baselines only. Cross-matching 
against Wide-field Infrared Survey Explorer (WISE) infra-red
catalogue of point sources \citep{r:wise} revealed 
WISE J015715.32$+$315419.2 object with magnitude 13.9 at 
3.6~$\mu$ within $0''.36$ of 0154$+$31A.

\begin{figure}
   \includegraphics[width=0.40\textwidth]{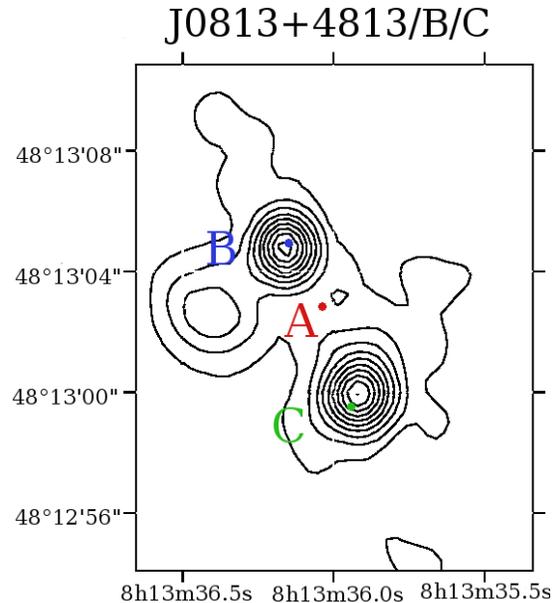}
   \caption{Triple source 0809$+$483/0809$+$48B/0809$+$48C
            VLA image at 4.86~GHz on epoch 1995.10.15 with
            beam size $1''.27$, project AS561, is shown as 
            a contour map. VLBI positions are shown with filled 
            circles. Component~A is the most compact and has flux 
            density 5~mJy at intercontinental baselines. 
            Components~B and C are resolved and visible only at short
            baselines.
   }
   \label{f:J0813+4813}
\end{figure}

{\bf 0809$+$483/0809$+$48B/0809$+$48C} has three components 
within $6''$. Components~B and C coincide with extended radio-lobes 
visible at the VLA image (Figure~\ref{f:J0813+4813}). Compact 
component~A that has X-band flux density 5~mJy at intercontinental 
baselines is located  between radio lobes. WISE 
J081336.05$+$481302.7 with magnitude 13.5 at 3.6~$\mu$ was found 
within $0''.11$ of component~A. 

\begin{figure}
   \centerline{\includegraphics[width=0.48\textwidth]{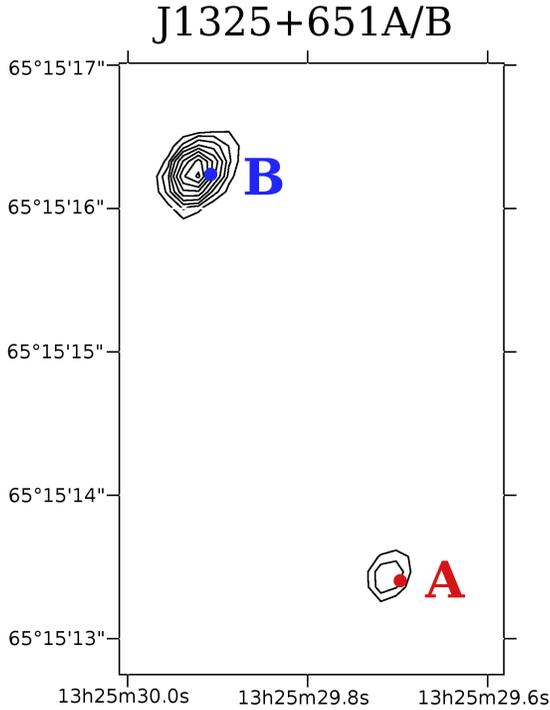}}
   \caption{Double source 1323$+$65A/1323$+$65B
   VLA image at 8.4~GHz on epoch 1999.08.08 with beam 
   FWHM $0.''27$, project AR415, is shown as a contour map. 
   Although component~B is stronger at the VLA image, it 
   is resolved out at X~band at VLBA resolution. Component~A
   has flux density 27~mJy at both bands.
   }
   \label{f:J1325+6515}
\end{figure}

{\bf 1323$+$65A/1323$+$65B} has two components with separation
$3''.157$. Although component~B is 5 times stronger at X~band 
at the VLA image (Figure~\ref{f:J1325+6515}) than component~A 
--- 21~mJy versus 4~mJy, it is not detected at VLBA scale at X~band. 
Component~A with correlated flux density 4~mJy at intercontinental
baselines lies within $0''.20$ of WISE J132529.70$+$651513.3
which has magnitude 14.6 at 3.6 $\mu$.

\begin{figure}
   \includegraphics[width=0.40\textwidth]{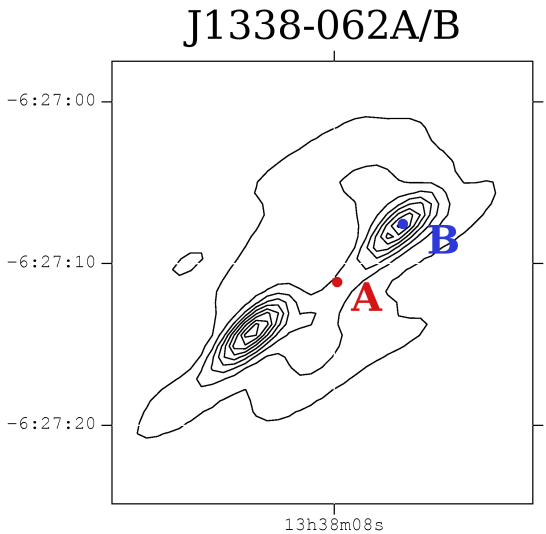}
   \caption{Double source 1323$+$65A/1323$+$65B at 
   the VLA image at 1.4~GHz on epoch 2001.04.29 with beam 
   size $6.''45$, project AB950 (FIRST), is shown as a contour map. 
   Component~B was detected in OBRS--2 only at S~band.
   }
   \label{f:J1338-0627}
\end{figure}

{\bf 1335$-$06A/1335$-$06B} has two components $4''.896$ apart.
Component~B, associated with a radiolobe (Figure~\ref{f:J1338-0627})
was detected at S~band only. Compact component~A with flux density
7~mJy at X~band is within of $0''.18$ of WISE J133807.98$-$062711.0,
which has magnitude 13.9 at 3.6 $\mu$.

\begin{figure}
   \includegraphics[width=0.40\textwidth]{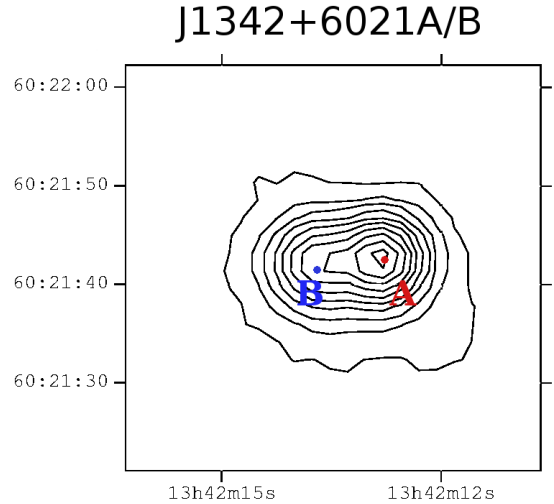}
   \caption{Double source 1340$+$60A/1340$+$60B
   VLA image at 1.4~GHz on epoch 2002.02.07 with beam 
   size $5.''4$, project AB950 (FIRST), is shown as a contour map. 
   Component~B was detected only at S~band.
   }
   \label{f:J1340+6021}
\end{figure}

{\bf 1340$+$60A/1340$+$60B} has two components separated at $3''.014$.
The source looks elongated at the VLA image at 1.4~GHz 
(Figure~\ref{f:J1340+6021}). Component~B was detected at S~band 
only. Component A is within $0''.23$ of WISE J134213.26$+$602142.9 
which has magnitude 14.2 at 3.6 $\mu$.

\section{Discussion}
\label{s:discussion}

   Position accuracy, 2--5~mas for 80\% OBRS--2 sources, is too coarse
to make a meaningful comparison with \Gaia\ because the frequency setup
was not favorable for precise absolute astrometry.

   Cross-referencing the cumulative catalogue of radio sources detected
with VLBI in the absolute astrometry mode at 8~GHz against the optical 
catalogue of active galaxy nuclea, including quasars, of \citet{r:vcv2010}, 
I found that 1676 objects, or 23\%, have a counterpart within a $4''$
search radius. Of them, 825 objects are brighter V~$18^m$ mag. Of them, 377, 
or 46\%, were observed in OBRS--1 or OBRS--2 programs, and 293 of them were 
observed only in these two programs. Five OBRS--2 sources, 0012$+$305, 
0232$-$042, 0744$+$092, 1146$+$249, 1632$+$198 have position offsets with 
respect to the optical catalogue of quasars of \citet{r:vcv2010} 
exceeding~$4''$. Since their offsets with respect to WISE catalogue are 
in a range of $0.08''$--$0.43''$, I consider that their positions in 
the optical catalogue had an error.

\begin{figure}
   \par\vspace{3ex}\par
   \includegraphics[width=0.48\textwidth]{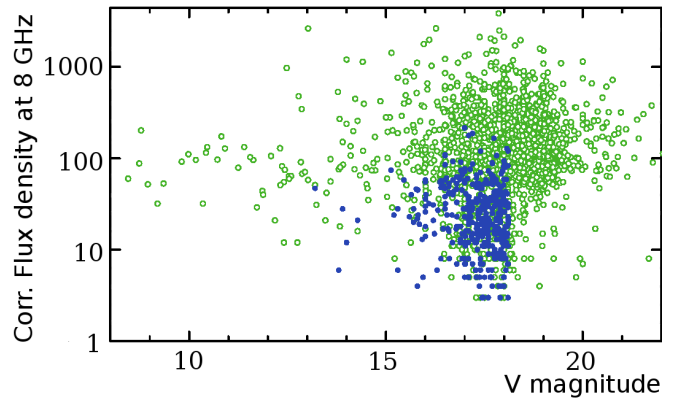}
   \caption{Dependence of the median correlated flux density at 8~GHz
            at baselines longer than 5000~km versus V magnitude 
            of those active galaxy nuclea that have been detected with
            VLBI under absolute astrometry programs. Solid blue filled
            circles show sources detected at X~band at OBRS--1 and 
            OBRS--2 campaigns. Green hollow circles show associated 
            sources observed under other programs.
   }
   \label{f:v_flux}
\end{figure}

  Figure~\ref{f:v_flux} shows the dependence of the correlated flux 
density versus V magnitude for the entire sample of radio-optic
associations and for the sub-sample observed in the OBRS--1 and OBRS--2
programs. There is no sign of obvious correlation between optical
brightness and radio brightness. We see that the OBRS--1/OBRS--2 sources 
are systematically weaker than those observed in other programs. 
Of 400 sources detected in OBRS--1 and OBRS--2, only 148 have an unresolved 
component at X~band stronger than 30~mJy, while the total number of radio 
sources associated with quasars brighter V~$18^m$ mag and with unresolved
component stronger than 30~mJy is 683. 

  If a source is too weak at long baselines, the position uncertainty 
0.1~mas will not be achieved for a reasonable integration time because 
of the thermal noise. The median semi-major axis of the position error 
ellipse of OBRS--1/OBRS--2 sources without reweighting is 2.2~mas. 
The frequency sequence used in these observing programs results 
in a group delay uncertainty at a given SNR that is a factor of 11.2 
greater than the group delay uncertainty of regular geodetic observations 
under RDV program. Even if OBRS programs were observed with the same 
frequency sequence as RDV experiments, the median position uncertainty 
due to the thermal noise would have been 0.2~mas. Therefore, future 
observations for improving source positions associated with optically 
bright quasars should be focused on observing compact radio sources 
with a strong unresolved core. {\it The majority} of such objects 
were detected in program {\it other} than OBRS--1/OBRS--2. Therefore, 
program OBRS can be considered as partially successful. Observing sources 
known as weak from the EVN detection survey \citep{r:bou10} was not 
warranted for the goal of the project. Selection of a frequency 
sequence that is unfavorable for astrometry degraded position accuracy 
by one order of magnitude but did not bring any merit.

  The VLBI catalogue is complete only to flux densities 
180~mJy \citep{r:aofus}. Figure~\ref{f:v_flux} suggests there may exist 
other strong radio sources associated with optically bright quasars.
Systematic surveys targeted to sources with correlated flux densities 
at long baselines in a range of 50--180~mJy promise to reveal new radio 
loud quasars. If to observe each target source for 2 minutes at 512~Mbit/s
at X/S in two scans each at the VLBA, $\sim\!1800$ sources with correlated
flux densities down to 25~mJy could be observed for 242 hours allotted 
for OBRS--1, OBRS--2, and the EVN detection survey. This approach is 
an alternative to the strategy adopted for OBRS project. 

\section{Summary}
\label{s:summ}

   Analysis of the second dual-band S/X VLBA campaign of the program for
observing optically bright extragalactic radio sources allowed me to
determine positions of 295 target sources and make images of 285 of them. 
Because of using the frequency setup unfavorable for absolute astrometry, 
the position uncertainties ranged from 2 to 200~mas with the median value 
of 3.2~mas. Many these sources are suitable as phase calibrators.

  This position accuracy is sufficient for using these sources as phase 
calibrators, but not sufficient for drawing meaningful conclusions from 
comparison of \Gaia\ and VLBI positions. Approximately 1/3 of observed
sources have strong unresolved core and their positions can be determined
with accuracies better than 0.1~mas in future VLBI observing programs with
appropriate frequency setup.

\acknowledgements

The National Radio Astronomy Observatory is a facility of the 
National Science Foundation operated under cooperative agreement 
by Associated Universities, Inc. This publication makes use 
of data products from the Wide-field Infrared Survey Explorer, 
which is a joint project of the University of California, and 
the JPL/California Institute of Technology, funded by the NASA.

{\it Facilities:} \facility{VLBA (project code GC034)}.

\end{document}